\documentstyle[aaspp4]{article}
 
\begin{document}

\title{Far Ultraviolet Performance of the Berkeley Spectrograph
During the ORFEUS-SPAS II Mission$^1$}

\author{Mark~Hurwitz, Stuart~Bowyer, Robert~Bristol,
W.~Van~Dyke~Dixon, Jean~Dupuis, Jerry~Edelstein, Patrick~Jelinsky, 
Timothy~P.~Sasseen$^2$, and Oswald~Siegmund}
\affil{Space Sciences Laboratory,
University of California,\\ Berkeley, CA 94720-7450}
\altaffiltext{1}{ Based on the development and utilization of ORFEUS
(Orbiting and Retrievable Far and Extreme Ultraviolet Spectrometers),
a collaboration of the Institute for Astronomy
and Astrophysics of the University of Tuebingen, 
the Space Astrophysics Group, University of California,
Berkeley, and the Landessternwarte Heidelberg.}
\altaffiltext{2}{currently with Department of Physics,
University of California, Santa Barbara, CA 93106}

\authoremail{markh@ssl.berkeley.edu}

\begin{abstract}

The Berkeley spectrograph aboard the ORFEUS telescope made its second
flight on the 14-day ORFEUS-SPAS II mission of the Space Shuttle {\it Columbia}
in November/December 1996.  Approximately half of the
available observing time was dedicated to the Berkeley spectrograph,
which was used by both Principal and Guest Investigators.  The
spectrograph's full bandpass is 390--1218 \AA; here we discuss its
in-flight performance at far-ultraviolet (FUV) wavelengths, where most
of the observations were performed.  The instrument's effective area
peaks at 8.9 $\pm$ 0.5 cm$^2$ near 1020 \AA, and the mean spectral
resolution is 95 km s$^{-1}$ FWHM for point sources.  Over most of the
spectral range, the typical night-time background event rate in each
spectral resolution element was about 0.003 s$^{-1}$.  Simultaneous
background observations of an adjacent blank field were provided
through a secondary, off-axis aperture.  The Berkeley spectrograph's
unique combination of sensitivity and resolution provided valuable
observations of approximately 105 distinct astronomical targets,
ranging in distance from the earth's own moon to some of the brightest
AGN.

\end{abstract}


\section{INTRODUCTION}

The German spacecraft Astro-SPAS, a deployable platform
designed to meet the technical performance demands of 
astronomical payloads and similar scientific instruments,
comprised the primary payload aboard shuttle mission STS-80 ({\it Columbia}).
On this, its third flight,
the platform carried a trio of far-ultraviolet instruments:
two independent spectrographs within
the 1 meter diameter ORFEUS telescope (Grewing et al. 1991)
and the IMAPS objective-grating spectrograph (Jenkins et al. 1996).
All three had flown on the Astro-SPAS' 5-day maiden voyage
in September of 1993,
but improvements in instrument performance, and the critical
need for additional observation time, motivated a reflight.
A photograph of the payload is shown in Plate~1.

With few exceptions, targets suitable for ORFEUS were too faint 
for IMAPS, so no attempt was made to coalign these 
instruments closely.  Within the ORFEUS telescope
a flip mirror was employed to direct the optical beam to
one spectrograph or the other.
Hence in general only one of the three instruments was operated at a time.
The available observing time
was shared equally between Guest Investigators selected by 
peer review and the Principal Investigator teams who had 
provided the instruments.
Flight operations were directed from a control complex  
at the Kennedy Space Center.

The general design of the Berkeley spectrograph
has been discussed previously (Hurwitz \& Bowyer 1986, 1996).
We changed the instrument between missions only by
overcoating of two of the four diffraction
gratings (including the far ultraviolet grating) with silicon carbide, 
introducing the multiple apertures discussed below, and 
modifying the detector electronics to improve the imaging 
at high count rates.
We did not recoat the KBr photocathode on the microchannel plate
detectors; the delay-line anode detector systems are 
discussed in Stock et al. (1993).
In this work we report on the performance and calibration of the 
spectrograph during the ORFEUS-SPAS II mission
and the instrumental effects of interest to Guest Investigators 
and other users of the extracted data products.

\section{THE ASTRO-SPAS PLATFORM AND ORFEUS TELESCOPE}

The Astro-SPAS was fabricated by Daimler-Benz Aerospace
in Ottobrunn, Germany.  An extremely reliable platform offering
high scientific performance at comparatively low cost,
it relies on nonrenewable resources such as cold gas thrusters,
batteries, and on-board recorders for primary data storage.
The spacecraft was deployed on 20 November, and recovered on 4 December, 1996.

When observing with the Berkeley spectrograph, the on-board 
recorders capture data at a rate of about 131 kbits s$^{-1}$.
Each photon event processed by our electronics
requires 24 bits of encoding: 8 bits of Y 
coordinate, 15 of X, and 1 for detector identification.
With an allowance for ancillary data, the maximum recordable
event rate corresponds to about 4400 spectral 
photons s$^{-1}$.  A slower telemetry link (8 kbits s$^{-1}$ for the scientific
instruments) enables a subset of the data to be transmitted
to ground via the shuttle except during
comparatively short periods (minutes to, on occasion, hours)
when prohibited by the orientation of the shuttle or
the unavailability of TDRSS.

For fine pointing, the spacecraft's attitude control system (ACS) 
utilizes tracking signals from a star tracker mounted parallel 
to the ORFEUS telescope axis (hereafter Z axis). 
Under normal circumstances the ACS achieves
an absolute pointing error less than about 5\arcsec\
and, during the ORFEUS-SPAS II mission, a jitter
of about $\pm$ 2\arcsec\ peak-to-peak.
The roll angle about the Z axis is determined automatically
by the ACS, which orients the telescope door for use as a sun shade.
If a guide star pattern is not recognized by
the tracker upon arrival at a specified target, the ACS is forced to
rely on gyroscopes, introducing an absolute pointing error
that generally places targets outside the 26\arcsec\ diameter 
spectrograph entrance aperture.
Outside the nominal 50~\arcdeg\ (half-angle) sun avoidance cone, 
lack of suitable guide stars caused the loss of one or two targets
during the ORFEUS-II mission.
Observations within this cone were possible only in fortuitous
circumstances (the presence of unusually bright guide stars
and/or occultation of the sun by the earth's limb).

The ORFEUS telescope systems performed well throughout the
mission.  The mechanisms of potential concern
(main telescope door, spectrograph flip mirror,
and diaphragm selection blade) operated nominally.
There are three available diaphragm positions, each of 
which corresponds to a unique aperture or combination 
of apertures at the focal plane.
Position 1 contains
a single on-axis aperture 20\arcsec\ in diameter
and was rarely employed for Berkeley observations.
Position 2 contains a single on-axis aperture 10\arcmin\ in diameter,
used during the initial star tracker/telescope coalignment
and for some observations of diffuse backgrounds and extended objects.
Position 3, most commonly used with our spectrograph,
contains three apertures.  For most
observations the target was placed in the on-axis, 26\arcsec\ diameter hole.
A second clear hole, about 1.4 times larger in area
and displaced by 2\arcmin .4, usually admitted only diffuse sky glow,
although serendipitous spectroscopy of astrophysical sources was performed
in some crowded fields. 
A third aperture, 120\arcsec\ in diameter,
is displaced by 5\arcmin .0 and covered by 
a tin filter approximately 1500 \AA\ thick.
The tin filter is virtually opaque to far-ultraviolet
radiation and was employed primarily for observation
of the EUV spectrum of the bright B star $\epsilon$ CMa.

\section{FLIGHT ACTIVITIES AND THERMAL EFFECTS}

A modest anomaly early in the mission delayed the initiation
of science observations by several hours.
Although preflight knowledge of the coalignment between the telescope 
and the star tracker is specified within $\pm$5\arcmin,
the actual misalignment (30 \arcmin) greatly exceeded the 
radius of the largest aperture.  This offset
caused some consternation among all the science teams until it 
was identified and resolved by the Berkeley group, which achieved final
on-orbit coalignment to within the few arcsecond level around 326/08:00 
(Day of Year / HH:MM, GMT).

Unlike sounding rockets and long-duration missions,
the time scale for thermal changes in the instrument
structure is neither much longer nor much shorter than the 
mission length.  By necessity, both ORFEUS spectrograph teams 
refined the coalignment several times during the first few days;
we also carried out frequent observations of the bright
symbiotic binary RR Tel to characterize the Berkeley spectrograph 
wavelength scale.  Based on thermal models and data from 
the ORFEUS-I mission, the structure should have cooled to near-equilibrium
conditions by day four or five.

Beginning about four days into the mission, 
the Astro-SPAS was twice required to assume a ``minimum drag''
configuration, in which the platform orientation was held fixed 
with respect to the instantaneous velocity vector.
These orientations were specified by shuttle flight directors 
to ensure a safe separation between the three
bodies that were co-orbiting at the time (ORFEUS-SPAS, the shuttle, 
and the Wake Shield Facility [a secondary deployed payload]).
Observations were not possible during these periods,
the first of which began around 329/04:00 and lasted
for $\sim$ 4 hours.  The IMAPS instrument was
active immediately after this period;
Berkeley operations resumed at 329/13:00.
No anomalies in the science instruments associated with the 
first minimum drag orientation have been noted.
The second minimum drag session began around 329/20:00,
lasting $\sim$ 12 hours.
The spacecraft orientation seems to have differed from the 
first period (there are multiple orientations
in which a minimum cross-section is presented).
Although the full effects were not recognized at the time,
the second minimum drag period subjected the spacecraft
to prolonged asymmetrical radiative thermal 
loads unlike those present during normal operations.
Temperatures increased significantly.
A large offset arose between the telescope and 
star tracker axes, causing the outright loss of some Echelle
spectrograph targets when observations were first resumed
and necessitating frequent refinement of the alignment matrix
as structures cooled. 
Berkeley observations resumed around 330/20:00. 
Post-flight analysis has revealed changes in the 
wavelength scale associated with thermal excursions
persisting for many hours after the end of the second minimum drag period.

Once thermal equilibrium was restored,
operations for the remainder of the mission
were fairly routine.  Essentially all of the lost 
observing time was recovered in an extra day appended 
to the preplanned time line.  The extra day 
was deeply appreciated by the science teams and 
guest investigators, as its approval required a careful
balancing of science return and mission risk.
A crew exit hatch had become stuck, making it impossible
(short of an emergency depressurization of the cabin)
for the astronauts to assist manually if difficulties
arose in the recovery of the platform.
Fortunately, recovery around 338/10:00 was routine.

\section{EFFECTIVE AREA AND BACKGROUNDS}

In-flight observations of the hot DA white dwarf HZ 43 provide the most
reliable flux calibration for the Berkeley spectrograph. To compute a
synthetic spectrum for HZ 43 we have adopted $T_{eff} = 50,000$ K and
$\log (g [{\rm cm}^{-2}]) = 8.0$, corresponding to the values adopted
by Bohlin et al. (1995) for the HST/FOS flux calibration. These values
agree with those in the literature (e.g., Napiwotzki et al. 1993).
We use the grid of LTE-blanketed pure hydrogen white dwarf synthetic
spectra computed by Vennes (1992) and used in the modeling of the
ORFEUS-I spectra of MCT 0455$-$2812 and G191$-$B2B (Vennes et al. 1996).
We normalize the model by the visual magnitude of 12.914 (Bohlin et al.
1995). We take into account the H I interstellar opacity, adopting a
column of $8.7 \times 10^{17} \rm cm^{-2}$ determined by Dupuis et al.
(1995). Interstellar absorption is important in the core of the strong
Lyman lines and near the convergence of the series.  An effective area
curve is derived by dividing the count rate spectrum of HZ 43 by the
synthetic spectrum (after convolution with the instrumental resolution
profile). The result is fit with a fifth-order Legendre polynomial to
retain the overall shape of the curve while removing structure on small
spatial scales.

Two early observations of HZ 43 were lost 
because of the alignment difficulties mentioned previously.
We use the helium-rich white dwarf MCT 0501$-$2858 
(observed at 326/12:20 and several times subsequently) as 
a transfer standard to define the effective area curve
during the first observing period.
The complete family of effective area curves is shown in Figure 1.

After 329/03:15, the effective area is smooth and,
aside from a possible change at the longest wavelengths, temporally stable.  
Most science observations were performed during the stable period.
At earlier times the instrument response exhibits
temporal variability.  The time-line presented in Table 1 is the simplest 
that is consistent with the available data and our understanding of 
the instrument. 

%\begin{center}
\begin{table}[htbp]
\caption{Early Mission Time-Line}
\begin{tabular}{|l|l|l|} \hline

Time & Event(s) & Result \\ \hline \hline

Before launch&MCPs adsorb gas, lose ``scrub''&Drop in Aeff near 1170 \AA \\ \hline

326/09:30 & Begin Berkeley Shift 2 & First successful science observations \\ \hline
326/12:20 & MCT 0501 observed & 1170 \AA\ dip present \\ \hline
326/19:20 & End Shift 2 & Use MCT 0501 curve for all Shift 2 obs. \\ \hline

After 326/19:20, & Optical contamination, source unk. & Overall lowering of Aeff\\
Before 327/06:20  & Some desorption of MCP gases? & \\  \hline
327/06:20 & Begin Shift 3 & High count-rate target completes ``scrub'' \\  \hline
327/06:59 & HZ 43 observed & 1170 \AA\ dip gone, overall Aeff lower \\ \hline
327/16:40 & End Shift 3 & Use first HZ 43 curve for all Shift 3 \\ \hline

328/17:03 & Begin Shift 4 &  \\ \hline
329/03:15 & HZ 43 observed & Aeff back to high level, no dip, stable \\ \hline
329/03:53 & End Shift 4 & Use mean of Shift 4 and subsequent  \\
 & & HZ 43 obs. for remainder of mission \\ \hline

\end{tabular}
\vspace{0.5in}
\end{table}

Because the effective area at early times (primarily during Shifts 2 and 3)
relies on our best understanding of a changing instrument response,
the flux calibration for these observations carries an uncertainty
that varies with wavelength and must be inferred from Figure 1.
After stabilizing, the effective area peaks at about 8.9 cm$^2$ at 1020 \AA.
Relative variations among curves during the stable period
correspond to no more than about $\pm$ 4\%.
We have estimated flux errors associated with uncertainties in the 
atmospheric parameters of
HZ~43 to be about 5\% by perturbing T$_{\rm{eff}}$
by $\pm$ 2000~K and the gravity by $\pm$ 0.2 dex.
We carried out a consistency check with G191B2B by applying our
effective area curve and comparing with synthetic spectra.
The agreement is only fair if we model the spectrum
of G191B2B with a pure hydrogen composition 
and atmospheric parameters from the literature.
Metals contaminate the spectrum of G191B2B (Vennes et al. 1996),
and a better agreement can be obtained by lowering the
assumed effective temperature.  We will investigate these
effects in more detail in a future paper using 
models including line blanketing by heavy elements.
We note that flux calibration for the ORFEUS-I spectra
relied on observations of G191B2B and a pure hydrogen
photospheric model.  Intercomparison of targets common to
both missions indicates that ORFEUS-I far ultraviolet
fluxes (final extraction of October 1995) are systematically
high, and should be multiplied by 0.9 .

When the spectral photon event rate approaches the 
capacity of the telemetry system, instrument deadtime
effects become important.  The detector electronics
create ``stimulation'' events at the detector edges
at a rate (approximately 2 s$^{-1}$) that can be measured during
slews and other quiet periods.  The observed rate of these
events during on-target pointings
enables us to estimate the deadtime for
genuine spectral photons.  Data buffering and the
nonperiodic nature of the telemetry sampling system
ensure that the throughput for both stimulation
and actual photon events is identical.

At a given detector X (dispersion direction) coordinate,
the FWHM of the spectrum averages 2 to 3 pixels in Y.
Our extraction procedure sums the counts over
a 9-pixel (best-fit center $\pm$ 4 pixels) range,
which usually encompasses $>$ 96\% of the dispersed photon 
events.  One X-pixel corresponds to about 0.3 \arcsec\ of sky; 
one Y-pixel corresponds to about 10\arcsec.
The general background is scaled from 
two strips, each spanning 3 pixels in Y, immediately above
and below the spectral region of the detector.  

When the current flow parallel to the MCP surface cannot
replenish the electrons being drained from the channel walls,
the modal gain may sag.  Gain sag is a function of local count
rate caused by unusually bright emission-line or continuum targets.
Gain sag must therefore be discussed independently from overall deadtime effects,
which are a function of total count rate (including background).
Gain sag is of greatest concern near the center of the band, where
the astigmatism or spectral height is minimized.

In the Y-dimension, gain sag causes a loss of spatial
resolution, with the result that some source events are not
contained within the nominal 9 Y-pixel extraction window.
(Spectral spill-over unrelated to gain sag may also occur
if the source falls very near the edge of the entrance aperture.)
The remedy is to study the background extractions for
evidence of spectral spill-over,
and to adjust the background subtraction procedure accordingly.

The delay-line system that calculates the photon X coordinates is
generally robust against gain sag, but high local count rates can 
lead to wavelength-dependent variations in throughput.
A very bright continuum source may suffer an uncorrected
loss of throughput over broad regions where the astigmatism is small.
In addition, very local variations in the modal gain
or in the rate of electron replenishment
can mimic spectral features when the modal gain is severely depressed.
Several bright stars, substantially exceeding the recommended count rate limit,
were observed to allow intercomparison of our spectrograph
with the less sensitive Echelle instrument.
For one such star, with a flux of 1.5 $\times 10^{-10}$ ergs cm$^{-2}$ \AA$^{-1}$ s$^{-1}$ 
at 1050 \AA, gain sag creates a 33\% loss in throughput 
from roughly 1040 to 1060 \AA.  Spectral artifacts are also present
in this region.  These effects will be studied in greater detail,
but are of concern for only the brightest targets.

At wavelengths
longward of 1017 \AA, the background event
rate summed over 9 pixels in Y and 0.33 \AA\ in
wavelength (e.g., one spectral resolution element) 
is about 0.0044 s$^{-1}$ during the day and 0.003 s$^{-1}$ at night.
Most of the night-time background is attributable to
the detector dark rate and cosmic rays; the day-time increase is probably
associated with stray Lyman $\alpha$ radiation.
At wavelengths below about 976 \AA, the background is
significantly higher.  This region of the detector is exposed to
a light leak qualitatively different from the general stray light
affecting longer wavelengths.  This ``bright corner'' of the detector
was noted during the ORFEUS-I mission.  Baffles added to the
optical path were expected to attenuate the leak to negligible levels,
but did not accomplish this goal.  Our current
hypothesis is that radiation reaches these corners of the
detector via the spectrograph flip mirror in its retracted position.
In the ``bright corner,'' the corresponding day and
night rates are 0.091 and 0.012 s$^{-1}$.
Their ratio corresponds roughly to the day/night ratio 
of diffuse Ly $\alpha$.

The irregular footprint of the light leak 
creates a crossover region between about 976 and 1017 \AA\,
within which the background cannot simply be scaled from the
strips above and below the spectrum.
To estimate the background here, we assume that the
shape of the background in the spectral region 
and in the adjacent strips differs only 
by a translation in the X coordinate, which we
determine from detector images collected during slew periods.
It is possible for this procedure to introduce
artificial features within the overlap region.
Artifacts of this type are not expected to be as narrow
as unresolved spectral features, nor are they
likely to exceed 10\% of the background in amplitude.
As such, they are of potential concern only for certain
types of analysis of very faint targets.
Over much of the overlap region the bright O I airglow
complex near 989 \AA\ contaminates the spectrum in any case.

Grating scatter within the plane of dispersion
creates a false continuum that is not eliminated
by subtraction of the general background.
Grating scatter will be roughly constant in
counts per unit wavelength.
Its magnitude can be estimated from portions of the spectrum where broad 
absorption lines or ionization edges extinguish the astrophysical flux;
in our investigations of early-type stellar and white dwarf spectra
we have found that grating scatter contributes about 0.28\% of the 
average flux across the FUV band.
The rapid decline in photocathode sensitivity with
increasing wavelength ensures that grating scatter will
not greatly exceed this level even for sources whose
spectra rise fairly steeply.

Diffuse emission lines of local (geocoronal and interplanetary) origin
are superimposed on the source spectrum.  A spectrum obtained through
the off-axis aperture can be used to estimate the strength of these
lines.  Care must be taken, however, to account properly for the
larger diameter of the off-axis aperture, which increases not only
the total counts but the width of the point-spread function for 
diffuse emission.
To estimate the counts in the on-axis spectrum, the corresponding
counts in the off-axis spectrum must be divided by 1.4 $\pm$ 0.12. The
uncertainty in the scaling factor reflects small-scale nonuniformities
in the detector and imposes a limit on the reliability of the
subtraction that becomes important when the brightness of the diffuse
emission greatly exceeds that of the astrophysical source.
The off-axis spectrum for each science observation,
on-axis spectra of blank fields, and an IDL software procedure 
are available to facilitate the subtraction of line profiles.
Finally, it is evident that diffuse Lyman $\alpha$ 
affected the microchannel plate gain and hence
the effective area (at the $\sim$ 10\% level)
during the mission.  The sensitivity loss will 
be time-dependent, and may affect the on- and
off- axis spectra differently.  There are few
sources, however, bright enough to permit meaningful
measurements of the astrophysical flux at the
wavelength of the bright diffuse Lyman $\alpha$ line.

\section{SPECTRAL RESOLUTION AND WAVELENGTH SCALE}

To determine the instrument's monochromatic point 
spread function, we have fit the observed line profiles
of features in the spectrum of the bright symbiotic
binary RR Tel.  The profiles are well characterized
by Gaussian functions, with FWHMs generally increasing with
the ionization potential of the species
as found by Penston et al. (1983).
Correcting for the intrinsic width of the
lines (estimated from Penston et al.)
we find a mean instrumental profile of about 95 km s$^{-1}$ FWHM
or about 0.33 \AA.

Pre-flight laboratory measurements yielded a spectral 
resolution that is higher than the in-flight value 
by a factor of about 1.5.  During the ORFEUS-SPAS I mission
operational and scheduling difficulties had interfered with 
focusing activities, leading us to attribute the resolution 
loss to an uncorrected defocus.  Early in the second
mission we successfully observed RR Tel at a range of 
grating focus positions, but the line widths did not 
sharpen as much as expected.  The detector imaging performance
appears nominal; the aforementioned stimulation events
are as narrow on orbit as in the laboratory.
The spacecraft jitter, well characterized by the
star tracker and, during the first mission, by the 
Berkeley Fine Guidance detector, makes a negligible contribution.
It may be that some aspect of our laboratory setup --  
most probably a slight underillumination of the grating pupil -- 
led to an overestimate of the system performance.

The fairly wide scatter in the measured FWHMs and the
somewhat uncertain contribution of the intrinsic widths
make it difficult to assess the potential for systematic
variations of the instrumental profile with wavelength.
Based on preflight measurements, it is likely that the 
FWHM is more constant in $\Delta \lambda$ than in velocity.
Locally, the instrumental FWHM for any given feature
can be affected (at the $\sim$ 10\% level) by small-scale 
distortions introduced by the microchannel plate fabrication 
process.

The relationship between the observed wavelength and 
detector X coordinate is straightforward over most
of the bandpass.  Optical raytracing and in-flight
measurements of emission lines in RR Tel (longward of 950 \AA)
and the Lyman absorption series in the white dwarfs
(below 950 \AA) confirm that over most of the spectral range the
wavelength is a linear function of X plus a smooth departure 
that does not exceed $\pm$ 0.13 \AA.
Near the edges of the detector the image becomes compressed,
leading to a steepening of the $\lambda$ vs. X relationship.
At the short-wavelength end, the converging Lyman absorption
line series in the white dwarfs provides reference
features at closely spaced wavelengths (Hurwitz \& Bowyer 1995; note that
in the ORFEUS-SPAS I mission the Lyman convergence was well
separated from the edge of the detector).  There is no corresponding
reference pattern at the long wavelength edge of the detector,
but longward of Lyman $\alpha$ the spectrum becomes unreliable
in any case because of a reflection near that detector edge.

Temporal changes in the wavelength scale include
both an offset and a linear ``stretch.''
We use a combination of diffuse emission lines
and the repeated observations of RR Tel to characterize 
these effects.  

The wavelength assigned by the current version of our
spectral extraction software (as of 9 September 1997)
reflects all the effects mentioned here, including 
the anomalous thermal conditions following the second
minimum drag orientation.
Relative wavelength errors within a 
given spectrum should be less than 0.1 \AA\ between 915
and 1216 \AA.  However, because of the unknown placement of each
target within the aperture, an overall offset
of $\pm$ 0.5 \AA\ may exist in any individual spectrum.

To establish a wavelength scale for the off-axis aperture, 
we apply a fixed offset, determined for an average of 
the strongest diffuse emission features, to the on-axis 
wavelength solution.  Because the detectable diffuse emission
features are widely spaced in wavelength, residual uncertainties
(potentially at the 0.1 \AA\ level) do not create real ambiguity
in their wavelength identification.

\section{DETECTOR FLAT-FIELD EFFECTS}

Prior to delivery of the instrument we characterized the 
detector's response to flux that is uniform (at least on small spatial scales) 
by directing a bright beam produced by an Argon gas discharge
at a bead-blasted metal surface near the diffraction gratings.
The beam flux is dominated by a handful of strong emission lines 
in the FUV bandpass. 
We collected 159,000 seconds of data 
with the far-ultraviolet detector operating at a moderate
count rate (2000 to 4000~s$^{-1}$).

The flat-field illumination strikes the
detector at approximately the same mean angle
as does the genuine spectrum, but the
spectrum is the convergence of a fast f/5 beam,
whereas the flat-field illumination
diverges from a small spot about 1 meter from the detector.
The laboratory flat-field was collected under thermal 
conditions different from those on orbit. 
The detector response may be affected by
the total or local count rate, mechanical shifts caused
by launch (or landing) loads, or the simple passage of time.  
The laboratory flat-field is therefore useful for
characterizing the distribution of fluctuations in
the detector response, but not for detailed division 
of flight spectra.  

In flight observations, detector artifacts
may affect the dispersed astrophysical flux,
the background that is present within the spectral extraction window
(the contributing background), or the
background that is measured outside the extraction window 
(the subtracted background).
To estimate the magnitude of the first effect, we
extracted a ``spectrum'' from the preflight flat-field
image using the best-fit Y centroids derived from
the flight data, weighting the counts in
adjoining Y pixels in a fashion that closely
approximates the distribution of events 
in actual flight spectra.
To estimate the second effect, we performed
a similar extraction, adopting uniform
weighting over the full 9 Y-pixel spectral window.

From the resulting extractions we can
determine the distribution of detector artifacts
appropriate for a given spectral wavelength interval.
We first bin the extractions at a given $\Delta \lambda$,
then calculate the RMS dispersion of the excursions
about an extraction that has been smoothed by 7 $\Delta \lambda$.  
The size of the smoothing interval is somewhat arbitrary.
A broader smoothing would be more conservative, 
but unrealistic for most applications.  In fitting an
absorption line, for example, one would normally 
interpolate a continuum based on nearby
adjacent spectral resolution elements.
We show the results in Figure 2, where we have 
expressed the RMS or 1$\sigma$ width 
of the distribution as a fraction of the measured signal.
Shot noise contributes negligibly to these results
and has been subtracted in quadrature.
The upper tracing (solid line) is applicable to
dispersed spectral counts; the lower tracing
(dashed line) is applicable to contributing background counts.
Below about 0.15 \AA\ the curves fall precipitously
because the finite detector resolution smoothes
out such small-scale fluctuations.
At the nominal instrument resolution of 0.33 \AA,
the RMS is about 6.4\% and 5.5\% of
signal for the spectral and contributing 
background events, respectively.
Binned at broader wavelength intervals, the
dispersion of detector artifacts falls fairly
rapidly.  Even at 3 \AA, however, RMS fluctuations exist 
at the level of 3 to 3.5\% of signal.
These fluctuations probably result from the microchannel 
plate boule structure; each boule spans about 3 \AA\ of spectrum.

Artifacts in the {\it subtracted} background are 
generally negligible.  In the Y dimension, such artifacts 
are automatically smoothed by 6 pixels (uniform weighting).  
Because the background varies only slowly in 
the X dimension, it can generally be smoothed
fairly broadly in wavelength.
For most applications we recommend smoothing
the subtracted background by a boxcar with 
a full width of 1.2 \AA.
Smoothing over a broader interval 
can introduce difficulties near the crossover
region where the background varies comparatively
rapidly with wavelength, but may be helpful elsewhere.
In the extracted spectral file, we pre-smooth
the background by a uniform boxcar with a 
comparatively narrow width (0.19 \AA).
This retains effectively 100\% of the system
spectral resolution, which is important if
a portion of the source spectrum has spilled
over into the background region.
It is straightforward for the user to introduce
the additional recommended background smoothing;
the IDL procedure we provide to read the spectrum 
smoothes the background spectrum automatically.

\section{SUMMARY / FUTURE PLANS}

The Berkeley spectrograph aboard the ORFEUS telescope
offers a unique and important combination of spectral
resolution and effective area in the comparatively
unexplored far-ultraviolet wavelength band.  
During the ORFEUS-SPAS II mission in November/December 1996,
Principal and Guest Investigators utilized the spectrograph 
to observe some 105 astronomical targets.
These data will enter the public domain in early to mid 1998.

Lyman/FUSE is not far from launch, and will offer
a much higher spectral resolution and sensitivity 
for point sources.
However, near-trivial modifications would 
enable the Berkeley spectrograph to achieve a significantly
superior performance for studies of extended emission.
New replicas of the current gratings and SiC overcoating
of the primary mirror would allow the spectrograph to 
achieve 2 \AA\ slit-limited resolution 
through a 45 $\times$ 420\arcsec\ slit
and an effective area of $\sim$ 50 cm$^2$ 
across the 900 -- 1200 \AA\ band.
The angular resolution along the slit length would be 
better than 1 \arcmin.
Such an instrument would be highly desirable for
studies of intracluster gas, galaxies and their halos, 
supernova remnants, and other extended objects.
At time of writing, there are no specific plans for a third 
flight of the payload.  

\acknowledgements

We thank the NASA and DARA personnel whose efforts
made the ORFEUS-SPAS II mission successful, especially
the talented and patient crew of STS-80.  
Konrad Moritz and the Astro-SPAS operations team
provided a marvelous spacecraft and a high level of 
support throughout the mission.
Our fellow scientists at Tuebingen and Heidelberg 
and partners at Kayser-Threde GmbH made the ORFEUS 
program a scientific success and a personal pleasure.
We thank Stephane Vennes for providing a grid of white dwarf model
spectra that helped to determine the effective area curve, and
offer special thanks to Brian Espey of the Johns Hopkins University
for his generous assistance with the wavelength calibration.
We acknowledge the support of NASA grant NAG5-696.

\clearpage

% FIG 1
% \figcaption{H\,{\sc i} 21 cm profiles from Stark et al.\
% for the four starburst sight lines of $(a)$ IRAS 08339+6517, $(b)$ Mrk 1267,
% $(c)$ Mrk 66, and $(d)$ Mrk 496.
% Vertical dashed lines indicate the velocity range
% within which we assume that the H\,{\sc i} detections are secure.}

% FIG 2
% \figcaption{Gas phase interstellar medium transmission
% curve calculated for the sight line toward IRAS 08339+6517.
% Theoretical transmission has been convolved with a 3~\AA\
% FWHM Gaussian and rebinned at 0.51 \AA.
% Only H~I and associated metal lines are included.}

%%%%%%%%%%%%%%%%%%%%%%%%%%%%%%%%%%%%%%%%%%%%%%%%%%%%%%%%%%%%%%%%%%
% TO INCOPORATE FIGURES INTO PAPER, UNCOMMENT THE LINES BELOW
% AND COMMENT OUT THE \figcaption LINES ABOVE.  aef 2/13/97

% FIG 1
\begin{figure}
\plotone{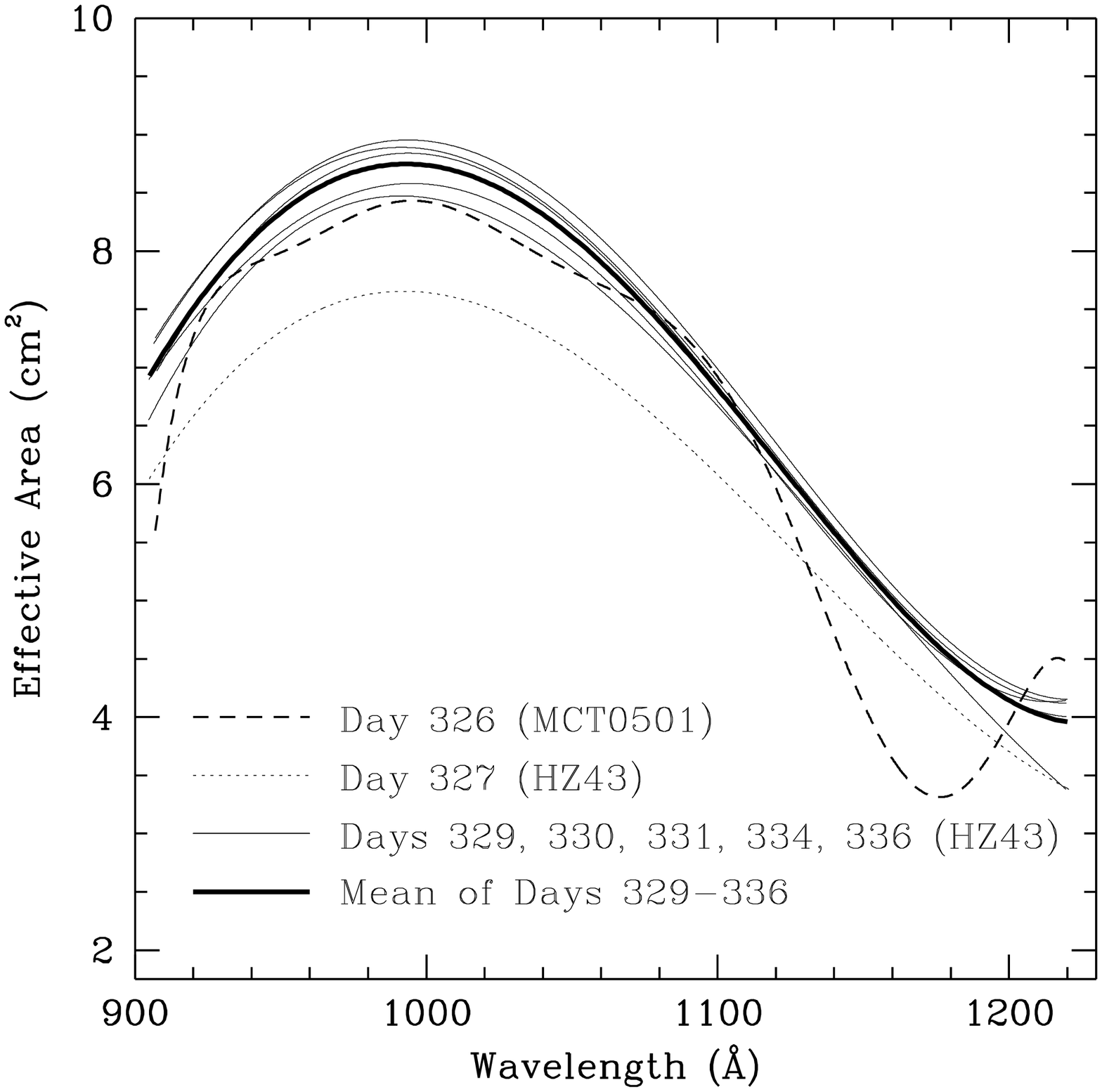}
\caption{Far ultraviolet effective area curve
for Berkeley spectrograph during ORFEUS-SPAS II mission.}
\end{figure}

% FIG 2
\begin{figure}
\plotone{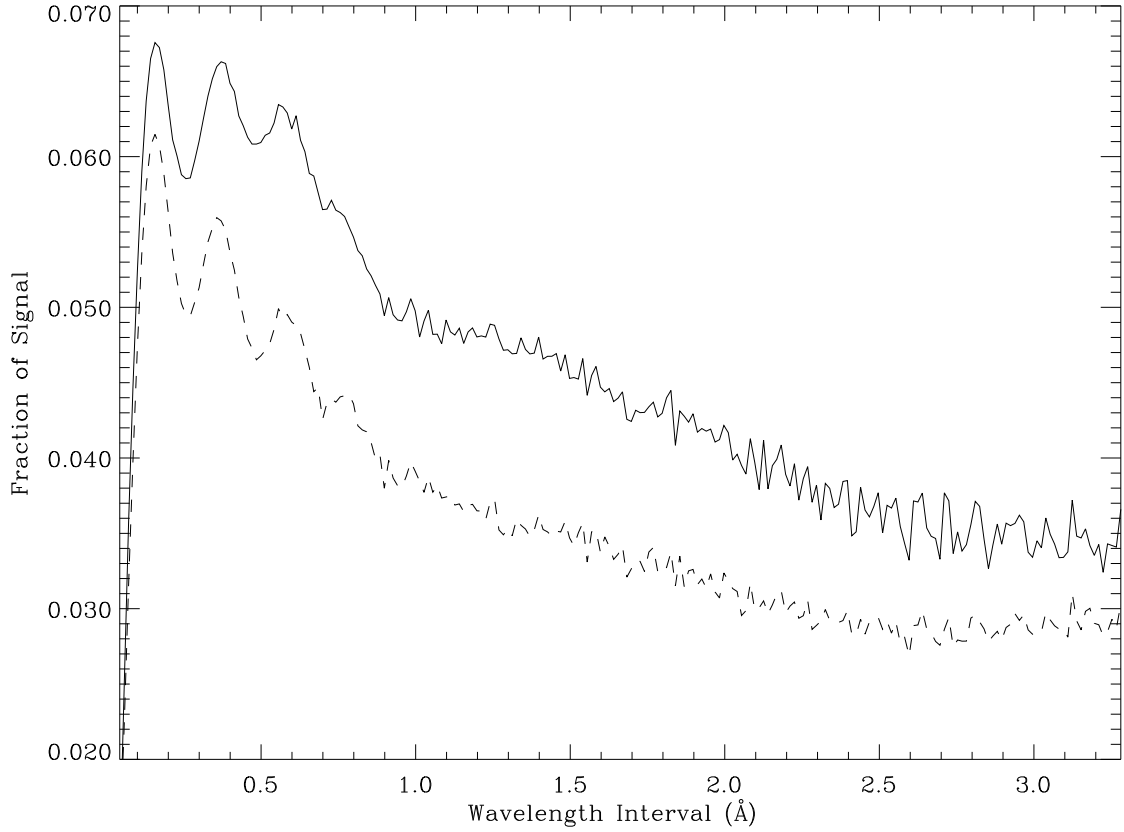}
\caption{RMS dispersion of detector artifacts expressed as
a fraction of the measured signal vs. wavelength interval.
These data are based on preflight illumination 
that is uniform on small spatial scales.  
The solid line applies to dispersed spectral photon events;
the dashed line applies to the contributing background events.
The rapid fall at small wavelength intervals results from the
finite detector resolution.  At large wavelength intervals, the function
declines because the increasing microchannel plate surface area leads 
to a smoother response.}
\end{figure}

%
%
% FIG 1
% \begin{figure}
% \plottwo{fig1a.ps}{fig1b.ps}
%  
% \vspace{.1in}
%  
% \plottwo{fig1c.ps}{fig1d.ps}
% \caption{H\,{\sc i} 21 cm profiles from Stark et al.\  
% for the four starburst sight lines of $(a)$ IRAS 08339+6517, $(b)$ Mrk 1267, 
% $(c)$ Mrk 66, and $(d)$ Mrk 496.
% Vertical dashed lines indicate the velocity range
% within which we assume that the H\,{\sc i} detections are secure.}
% \end{figure}
%
% FIG 2
% \begin{figure}
% \plotone{fig2.ps}
% \caption{Gas phase interstellar medium transmission
% curve calculated for the sight line toward IRAS 08339+6517.
% Theoretical transmission has been convolved with a 3~\AA\
% FWHM Gaussian and rebinned at 0.51 \AA.
% Only H~I and associated metal lines are included.}
% \end{figure}
%
%
%%%%%%%%%%%%%%%%%%%%%%%%%%%%%%%%%%%%%%%%%%%%%%%%%%%%%%%%%%%%%%%%%%

\end{document}